\documentclass[fleqn,usenatbib]{mnras}

\usepackage{newtxtext,newtxmath}

\usepackage[T1]{fontenc}
\usepackage{ae,aecompl}

\usepackage{graphicx}	
\usepackage{amsmath}	
\usepackage{amssymb}	

\newcommand{\beq}{\begin{equation}}
\newcommand{\eeq}{\end{equation}}
\newcommand{\beqn}{\begin{eqnarray}}
\newcommand{\eeqn}{\end{eqnarray}}
\newcommand{\grim}{{\tt grim}}

\title[Accretion disk with anisotropic viscosity]
{How important is non-ideal physics in simulations of sub-Eddington accretion onto spinning black holes?}

\author[F. Foucart et al.]{%
Francois Foucart$^{1}$\thanks{E-mail:
  fvfoucart@lbl.gov}\thanks{Einstein Fellow},
Mani Chandra$^{2}$,
Charles F. Gammie$^{3}$,
Eliot Quataert$^{4}$,
\newauthor
Alexander Tchekhovskoy$^{4,1,5}$\thanks{TAC Fellow}\\
$^{1}$Lawrence Berkeley National Laboratory, 1 Cyclotron Rd, Berkeley, CA 94720, USA\\
$^{2}$Research Division, Quazar Technologies, Sarvapriya Vihar, New Delhi, India, 110016\\
$^{3}$Department of Astronomy and Department of Physics, University of Illinois, 1002 West Green Street, Urbana, IL 61801\\
$^{4}$Departments of Astronomy and Physics and Theoretical
Astrophysics Center, University of California, Berkeley, CA, 94720\\
$^{5}$Center for Interdisciplinary Exploration \& Research in Astrophysics (CIERA),
Physics \& Astronomy, Northwestern University, Evanston, IL 60202, USA
}

\date{Accepted XXX. Received YYY; in original form ZZZ}

\pubyear{2015}

\begin{document}
\label{firstpage}
\pagerange{\pageref{firstpage}--\pageref{lastpage}}
\maketitle

\begin{abstract}
Black holes with accretion rates well below the Eddington rate are expected to be surrounded by low-density, hot, geometrically thick accretion disks. This includes the two black holes being imaged at sub-horizon resolution by the Event Horizon Telescope. In these disks, the mean free path for Coulomb interactions between charged particles is large, and the accreting matter is a nearly collisionless plasma. Despite this, numerical simulations have so far modeled these accretion flows using ideal magnetohydrodynamics.  Here, we present the first global, general relativistic, 3D simulations of accretion flows onto a Kerr black hole including the non-ideal effects most likely to affect the dynamics of the disk: the anisotropy between the pressure parallel and perpendicular to the magnetic field, and the heat flux along magnetic field lines. We show that for both standard and magnetically arrested disks, the pressure anisotropy is comparable to the magnetic pressure, while the heat flux remains dynamically unimportant. Despite this large pressure anisotropy, however, the time-averaged structure of the accretion flow is strikingly similar to that found in simulations treating the plasma as an ideal fluid. We argue that these similarities are largely due to the interchangeability of the viscous and magnetic shear stresses as long as the magnetic pressure is small compared to the gas pressure, and to the sub-dominant role of pressure/viscous effects in magnetically arrested disks. We conclude by highlighting outstanding questions in modeling the dynamics of low collisionality accretion flows.
\end{abstract}




\section{Introduction}
\label{sec:intro}

Supermassive black holes (SMBHs) are found at the center of most galaxies, and a majority of these black holes
are accreting much slower than the Eddington accretion rate~\citep{Ho2009}. This is in particular the case
for the two SMBHs with the largest angular size on the sky: SgrA$^*$, at the center of 
the Milky Way, and the black hole at the center of M87. These black holes
are important targets for high-resolution imaging experiments such as the Event Horizon 
Telescope~\citep{Doeleman2009} and Gravity~\citep{Eisenhauer2008}, which may provide us with constraints
on the properties of SMBHs and on the behavior of accretion flows in the immediate
neighborhood of a black hole. Accordingly, it is important to be able to quantitatively model 
accretion flows close to slowly accreting SMBHs.

Accretion flows at low accretion rates are expected to be geometrically thick and optically thin. Both
theoretical models~\citep{Yuan2014} and numerical simulations~\citep{Koide1999,DeVilliers2003,McKinney2004} 
of these {\it radiatively inefficient accretion flows} (RIAF)
indicate that the density and temperature in the disk are such that the mean free path for Coulomb interactions 
between charged particles is much larger than the typical size of the system, $\sim GM/c^2$ (with $G$ the gravitational
constant, $M$ the black hole mass, and $c$ the speed of light). We then expect the accreting matter to be nearly
collisionless. This raises questions about the accuracy of existing numerical simulations, which treat the accretion flow as an ideal fluid.

Shearing-box simulations using particle-in-cell methods have shown that the situation
is probably not as dire as it might seem. Wave-particle interactions due to velocity-space instabilities effectively increase
the collision rate of particles, and limit the amplitude of non-ideal effects~\citep{Kunz2014,Riquelme2015,Sironi2015,Kunz:2016}.
The impact of those instabilities is also verified observationally in the solar wind~\citep{Kasper2002,Hellinger2006}. More specifically,
the phase space of solar wind particles appears to be bounded by the instability thresholds of the mirror and firehose
instabilities, which limit the pressure anisotropy to be of the same order as the magnetic pressure. 

Global simulations
of an accretion disk made of a collisionless plasma would require time evolution of the 6D distribution 
functions of ions and electrons. This is beyond the reach of current numerical simulations. However, the effective
collisionality provided by wave-particle interactions justifies modeling the accretion flow as a {\it weakly collisional
plasma}, in which non-ideal effects are treated as a perturbation of the ideal fluid model.

Non-ideal effects could play a number of potentially important roles. First, they may affect the dynamics of the disk,
i.e. the bulk motion of the charged particles. Second, they may lead to different temperatures for the ions and
electrons. And finally, they may cause both ions and electrons to have non-thermal distribution functions.
The second and third points probably do not directly affect the dynamics of the disk, but may be important to its radiative
properties, which are largely determined by the energy spectrum of the electrons.

In this work, we focus on the impact of non-ideal effects on the dynamics of accretion disks at highly sub-Eddington accretion
rates. In particular, we include for the first
time in 3D global simulations the two non-ideal effects most likely to affect the disk dynamics: the
anisotropy between the pressure parallel and perpendicular to the magnetic field, and the presence of a heat flux along
magnetic field lines. To do this, we rely on our recently developed general relativistic model for the evolution of the pressure
anisotropy and heat flux in a weakly collisional plasma~\citep{Chandra:2015}. 

In previous axisymmetric global simulations of accretion
disks, we demonstrated that the pressure anisotropy rapidly grows and becomes comparable to the
magnetic pressure, saturating at the threshold of the mirror instability~\citep{Foucart:2016}. 
However, in axisymmetry, the turbulent cascade transfers
energy to longer wavelengths, and cannot be self-consistently sustained. To study a steady-state accretion flow,
we require the 3D simulations presented here. 

In Sec.~\ref{sec:emhd}, we review the
main features of the {\it extended magnetohydrodynamics} model used to describe the plasma in our simulations.
Sec.~\ref{sec:setup} discusses our choice of initial conditions and numerical methods. Sec.~\ref{sec:SANE} provides
a detailed comparison of simulations with and without non-ideal effects for an accretion disk in which the magnetic
pressure of the relativistic polar jet does not affect the dynamics of the disk (the {\it standard}, or {\it SANE} configuration),
Sec.~\ref{sec:MAD} shows results for a {\it magnetically arrested disk} ({\it MAD}), in which the magnetic pressure of the
jet nearly balances the pressure of the infalling gas. Secs.~\ref{sec:SANE}-\ref{sec:MAD} thus provide us with an opportunity
to study non-ideal effects in very different physical regimes. Finally, we conclude and discuss the current limitations of our
model and potential future work in Sec.~\ref{sec:conclusions}.

In the following sections, we use the length unit $r_g = GM/c^2 \sim 0.04 \left( \frac{M}{4\times10^6 M_\odot}\right)\,{\rm AU}$ and time unit 
$t_g = GM/c^3 \sim 20\left( \frac{M}{4\times 10^6 M_\odot}\right)\,{\rm s}$.

\section{Extended Magnetohydrodynamics model}
\label{sec:emhd}

We describe the nearly collisionless plasma as a {\it well-magnetized}, {\it weakly collisional} plasma,
following the general relativistic model proposed in~\cite{Chandra:2015}.
In this limit, we assume that the Larmor radius of charged particles is very small compared to the typical
length scale of the accretion disk ($L \sim r$, the distance to the center of the black hole). 
This is nearly certain to be the case in practice, as the disk is expected to be
unstable to the magnetorotational
instability~\citep[MRI,][]{Balbus1991}. Even a small seed magnetic field would grow exponentially
on a timescale comparable to the orbital time scale of the disk. 
We also assume that deviations from an ideal fluid
are small, so that non-ideal effects can be treated perturbatively. While not intuitively obvious, this assumption is
supported by particle-in-cell simulations showing that wave-particle interactions generated by velocity-space
instabilities (e.g. firehose, mirror) create an effective collision rate in the system~\citep{Kunz2014,Riquelme2015,Sironi2015,Kunz:2016}.
Under these assumptions, we model the plasma using an {\it extended magnetohydrodynamics} (EMHD) model
which is a generalization of Braginskii's theory for magnetized, weakly collisional plasmas~\citep{Braginskii1965},
modified to make the theory stable and causal within a general relativistic framework. 

We start from the ideal description of a fluid of density $\rho$, internal energy density $u$, pressure $P$,
and 4-velocity $u^\mu$, with a magnetic field 4-vector $b^\mu$ (in the fluid frame), and 
a known spacetime metric $g_{\mu \nu}$.
The ideal part of the stress-energy tensor 
is then
\beq
T^{\mu \nu}_{\rm ideal} = (\rho + u + b^2 + P) u^\mu u^\nu + (P+\frac{b^2}{2}) g^{\mu\nu} - b^\mu b^\nu,
\eeq
with $b^2 = b^\mu b_\mu$ being twice the magnetic pressure; we have absorbed the factor of $(4\pi)^{-1/2}$
into the definition of the magnetic field. Non ideal effects are modeled
through a heat flux constrained to be along magnetic field lines ($q^\mu = q b^\mu$), and an anisotropy between the pressure parallel and
perpendicular to the magnetic field lines, $P_\parallel = P - \frac{2}{3} \Delta P$ and $P_\perp = P + \frac{1}{3} \Delta P$. 
The full stress-energy tensor is then written as
\beq
T^{\mu\nu} = T^{\mu \nu}_{\rm ideal} + q^\mu u^\nu + q^\nu u^\mu + \Pi^{\mu\nu},
\eeq
where the anisotropic shear stress $\Pi^{\mu\nu}$ is related to the pressure anisotropy by
\beqn
\Pi^{\mu\nu} &=& -\Delta P \left( \hat b^\mu \hat b^\nu - \frac{1}{3} h^{\mu\nu} \right),\\
h^{\mu\nu} &=& g^{\mu \nu} + u^\mu u^\nu,
\eeqn
with $\hat b^\mu = b^\mu / b$.
In Braginskii's theory, the heat flux and pressure anisotropy are simply proportional to, respectively,
the temperature gradient along magnetic field lines, and the projection of the velocity shear tensor onto the magnetic field.
However, in general relativity, such a prescription violates causality. Instead, we have to promote $q$ and $\Delta P$ to evolved
variables, and drive them towards their desired value
\beqn
q_0 &=& \rho \chi \hat b^\mu \left(\nabla_\mu \Theta + \Theta u^\nu \nabla_\nu u_\mu \right),\\
\Delta P_0 &=& 3 \rho \nu \left( \hat b^\mu \hat b^\nu \nabla_\mu u_\nu - \frac{1}{3} \nabla_\mu u^\mu \right),
\eeqn
which are covariant versions of the heat flux and pressure anisotropy in Braginskii's theory.
Here, $\Theta = kT/(m_p c^2)$ is the dimensionless temperature, $m_p$ the proton mass,
$\chi$ the thermal diffusivity, and $\nu$ the kinematic viscosity. 

The first 8 evolution equations of the EMHD model are easily obtained from the conservation equations
\beqn
\nabla_\mu (\rho u^\mu) &=& 0, \label{eq:rhoev}\\
\nabla_\mu T^{\mu\nu} &=& 0 \label{eq:bianchi},
\eeqn
and from Maxwell's equation
\beq
\nabla_\mu (b^\mu u^\nu - b^\nu u^\mu)=0. \label{eq:induction}
\eeq
We also use the simple equation of state
\beq
P = \rho \Theta = (\Gamma-1) u,
\eeq
with $\Gamma$ the polytropic index of the fluid. In the simulations, we will consider an ideal gas
of non-relativistic particles, for which $\Gamma=5/3$. This choice is reasonable as long as the electron
temperature $T_e$ is much lower than the ion temperature $T_p$, and $\Theta_p<1$. This is expected to be the
case for slowly accreting supermassive black holes such as SgrA$^*$. 

For the non-ideal sector, we evolve the rescaled variables
\beqn
\tilde q &=& q \sqrt{\frac{\tau_R}{\rho \chi \Theta^2}},\\
\Delta \tilde P &=& \Delta P \sqrt{\frac{\tau_R}{\rho \nu \Theta}},
\eeqn
according to
\beqn
\nabla_\mu (\tilde q u^\mu) &=& -\frac{\tilde q - \tilde q_0}{\tau_R} + \frac{\tilde q}{2} \nabla_\mu u^\mu, \label{eq:evq}\\
\nabla_\mu (\Delta \tilde P u^\mu) &=& -\frac{\Delta \tilde P - \Delta \tilde P_0}{\tau_R} + \frac{\Delta \tilde P}{2} \nabla_\mu u^\mu \label{eq:evDP},
\eeqn
with $\tau_R$ a relaxation timescale yet to be determined. The exact choice of evolved variables and evolution equations
are driven in part by numerical considerations, and in part by the requirement that the model satisfies the second law of
thermodynamics. The system of equations defined by Eqs~(\ref{eq:rhoev},\ref{eq:bianchi},\ref{eq:induction},\ref{eq:evq},\ref{eq:evDP}) remains causal and stable as long as $\tau_R/\chi$ and $\tau_R/\nu$ are not too small,
in a sense discussed in more detail in~\cite{Chandra:2015}.

Note that the EMHD model used here includes an explicit treatment of
some quantities associated with dissipative processes (field aligned thermal diffusion
and viscosity) but relies on the numerical diffusion for the treatment of others (resistivity, cross-field viscosity and cross-field thermal diffusion). The motivation for this is that the transport along field lines is expected to be particularly large and important in low-collisionality plasmas.  It is worth bearing in mind, however, that ideal MHD simulations have demonstrated the importance of an explicit treatment of isotropic viscosity and resistivity for the saturation of the MRI in some regimes~\citep{Lesur:2007,Fromang:2007}. This is not included in our models but is an important extension to consider in future work (see Sec.~\ref{sec:future}).

This leaves us with 3 free model parameters: $\tau_R$, $\chi$, and $\nu$, which should be determined from kinetic theory.
 We can interpret $\tau_R$ as the effective mean-free time between collisions, which in our applications is the effective
 mean free time for wave-particle scatterings. In the absence of plasma instabilities, a reasonable first guess for $\tau_R$
 in an accretion torus is the dynamical timescale, $\tau_d =
 \sqrt{r^3/(GM)}$. We can also estimate $\chi$ and $\nu$ from
 their value in non-relativistic collisional theory, $\chi = \phi c_s^2 \tau_R$ and $\nu = \psi c_s^2 \tau_R$, with $\phi$ and $\psi$
 dimensionless constants of order unity and $c_s = \sqrt{\Gamma P / (\rho + \Gamma u)}$ the sound speed.
 Here, we choose $\psi=1$ and $\phi=1$.

 When $\Delta P$ becomes comparable to the magnetic pressure, however, the relaxation time $\tau_R$ is probably reduced
 by the onset of plasma instabilities. Particle-in-cell simulations~\citep{Kunz2014,Riquelme2015} have shown that the ions are 
 unstable to the firehose
 instability if $\Delta P < -b^2 \equiv \Delta P_{\rm firehose}$ and to the mirror instability if
 \beq
 \Delta P > \Delta P_{\rm mirror} = \frac{b^2}{2} \frac{P_\parallel}{P_\perp}.
 \eeq
 Once the pressure anisotropy reaches these thresholds, we expect the growth of plasma instabilities, and the saturation
 of $\Delta P$ at the instability threshold. Similarly, the heat flux cannot exceed the free-streaming value of $|q| = \rho c_s^3$. 
 In the EMHD model, we can attempt to mimic the effects of plasma instabilities by modifying $\tau_R$ in such a way that 
 $\Delta P$ remains within the expected stability region, and $|q|<\rho c_s^3$.
 This can be done if we choose, for example,
 \beq
 \tau_R = \tau_d \times (f_{\rm min} + f(|q|,\rho c_s^3) \times f(\Delta P,\Delta P_{\rm firehose}) \times f(\Delta P,\Delta P_{\rm mirror})
 \eeq
with
\beqn
f(x,x_{\rm max}) &=& \frac{1}{1+e^{g(x,x_{\rm max})}},\\
g(x,x_{\rm max}) &=& \frac{1}{\Delta x} \frac{x-x_{\rm max}}{x_{\rm max}}.
\eeqn
In our simulations, we use this prescription with $\Delta x = 0.01$ and $f_{\rm min}=10^{-5}$ (the exact form of the function $f$
sets how far $\Delta P$ and $q$ can grow above their maximum allowed value, and is chosen to provide a fairly strict limit in our model). 

When analyzing the results of our simulations, a few analytical properties of the EMHD model are good to keep in mind:
\begin{itemize}
\item For a Keplerian velocity profile and a plasma parameter $\beta = 2P/b^2 \gtrsim 1$, we expect $\Delta P \sim b^2/2$ and
$\tau_R \sim \tau_d/\beta$~\citep{Kunz2014}.%
\footnote{
$\tau_R \sim \tau_d/\beta$ is both the value of $\tau_R$ necessary to have $\Delta P \lesssim \Delta P_{\rm mirror}$ in the EMHD model, and the collision timescale measured in shearing-box
simulations of the mirror instability in the saturated regime~\citep{Kunz2014}. In this respect at least, our interpretation of $\tau_R$ as an effective
collision timescale is consistent with more accurate plasma simulations.
}
A priori, there is no guarantee that this is also true for a turbulent flow in a
magnetized disk. However, we have already shown that this prediction holds for turbulent accretion flows in axisymmetry~\citep{Foucart:2016},
and the same is true for the full 3D simulations presented here. Because of this, the arbitrary choices made for 
the values of $\phi$ and $\psi$ are relatively unimportant. Instead, the mirror instability threshold $\Delta P_{\rm mirror}\sim b^2/2$ sets
$\Delta P$. 
\item $\Delta P$ and $q$ are driven towards their Braginskii values on a timescale $\tau_R$ which is,
at most, comparable to the orbital time scale, and shorter than the time scale for the growth of the MRI.
This means that if the evolution of the magnetic field is mostly driven by the growth of the MRI, $\Delta P$ effectively reacts nearly
instantaneously to changes in the magnetic pressure. The time evolution equations for $\Delta P$ and $q$ are necessary
to guarantee causality, but are not otherwise critical.
\item The off-diagonal contributions of the viscous stress $\Pi^{\mu}_{\nu}$ and of the magnetic stress $T^{\mu}_{\nu,\rm mag}$
to the stress-energy tensor $T^{\mu}_{\nu}$ have some interesting similarities:
\beqn
T^{\mu}_{\nu, \rm mag} &=& -b^2 \left(\hat b^\mu \hat b_\nu - \frac{h^{\mu}_{\nu}}{2} - \frac{u^\mu u_\nu}{2}\right), \label{eq:Tmunumag}\\
\Pi^{\mu}_{\nu} &=& -\Delta P \left( \hat b^\mu \hat b_\nu - \frac{1}{3} h^{\mu}_{\nu} \right)\label{eq:Tmunuvis}.
\eeqn
For $\Delta P=b^2$, the off-diagonal components of the viscous and magnetic stresses as measured in the orthonormal frame
comoving with the fluid are equal.
In particular, this is true for the $r-\phi$ component of the stress-energy tensor in the comoving frame, which dictates 
angular momentum transport in the disk. On the other hand, the diagonal components of the magnetic and viscous stresses in the comoving
frame, which describe the energy density and isotropic pressure, differ.
\end{itemize}
These properties of the EMHD model will play an important role in the interpretation of our simulation
results in Sections~\ref{sec:SANE}-\ref{sec:MAD}.

\section{Numerical Setup}
\label{sec:setup}

\subsection{Initial Conditions}
\label{sec:ID}

For this first exploration of the effect of a pressure anisotropy and heat conduction on the dynamics of an accretion torus, we consider initial conditions
widely used in the literature: the hydrostatic equilibrium solution for a torus of constant
specific entropy orbiting a rotating black hole, due to~\cite{Fishbone1976}. 
Ideal MHD (IMHD) simulations of accretion tori have already shown that there are at least two distinct types of accretion onto 
black holes: the so-called SANE (Standard And Normal Evolution, \citealt{2012MNRAS.426.3241N}) and
MAD (Magnetically Arrested Disk, \citealt{1974Ap&SS..28...45B,1976Ap&SS..42..401B,2003PASJ...55L..69N,2003ApJ...592.1042I,2008ApJ...677..317I,2011MNRAS.418L..79T,2012MNRAS.423L..55T}) configurations. In a MAD disk, 
the magnetic flux threading the black hole horizon becomes so large that the magnetic pressure of the jet can temporarily stop the flow of matter into the 
black hole. In a SANE disk, on the other hand, the magnetically-dominated polar regions do not significantly affect the dynamics of the accretion flow.
Absent large-scale poloidal magnetic flux dynamo, magnetic flux
conservation implies that the magnetic field seeded in the initial
equilibrium torus determines the large-scale magnetic flux that can be accumulated by the black hole,
and thus the resulting type of the steady-state accretion flow. Here,
we will consider two sets of initial conditions, one leading to SANE
and one to MAD accretion flows.
Whether astrophysical accretion disks around supermassive black holes
are in a MAD or SANE state remains an open question. There are
observational indications that dynamically important magnetic fields
might be present in active galactic nuclei \citep{2014Natur.510..126Z,2014Natur.515..376G},
tidal disruption events \citep{2014MNRAS.437.2744T}, and
gamma-ray bursts \citep{2015MNRAS.447..327T}.

The Fishbone-Moncrief solution is entirely defined by the dimensionless spin of the black hole ($a/M_{\rm BH}$), the inner radius of the torus ($r_{\rm in}$) and
radius of the pressure maximum in the disk ($r_{\rm max}$), the polytropic index used in the equation of state $\Gamma=5/3$, and the entropy $P/\rho^\Gamma=0.0043$. For the SANE simulation, we choose $a/M_{\rm BH}=0.9375$, $r_{\rm in}=6r_g$, and $r_{\rm max}=12r_g$. For the MAD simulation, we have
$a/M_{\rm BH}=0.5$, $r_{\rm in}=15r_g$, and $r_{\rm max}=32r_g$. As the mass scale within the accreting matter is arbitrary, we choose it so that $\max{(\rho)}=1$ in the initial torus.

\begin{figure}
\includegraphics[width=\columnwidth]{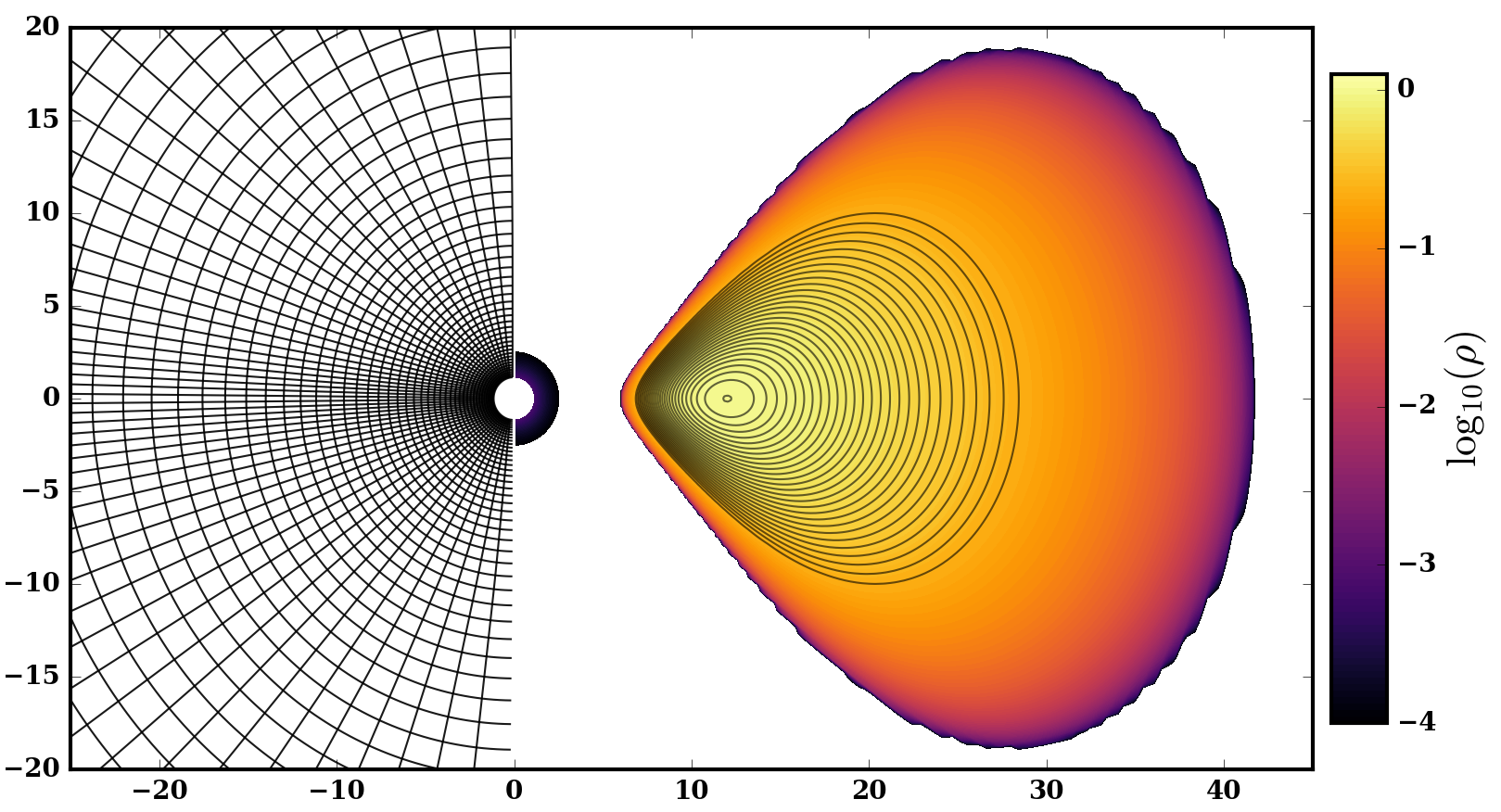}
\caption{Vertical slice through the numerical grid ({\it left}, with each cell actually representing a $5x5$ group of cells at our highest resolution),
and the initial torus for the SANE simulations ({\it right}, the color scale shows the fluid density, while solid black lines follow the seed magnetic field lines.).}
\label{fig:ID}
\end{figure}

We seed the tori with a weak poloidal magnetic field. The magnetic field in the inertial frame, $B^i$, is derived from the vector potential $A^\mu$ using 
$B^i = b^i u^t - b^t u^i = \epsilon^{ijk}\nabla_j A_k$, with $\epsilon_{ijk}$ the Levi-Civita symbol. For the SANE disk, we set $A_{\phi}=A_0 \max(\rho-0.2,0)$,
with the constant $A_0$ chosen so that the minimum value of $\beta = 2P/b^2$ in the disk is $\beta_{\rm min}=15$. This guarantees that the magnetic field is weak
enough to avoid disrupting the equilibrium torus, but strong enough to allow us to resolve the fastest growing mode of the magnetorotational instability (MRI). 
For the MAD disk, we set $A_{\phi}=A_0 \rho^2 r^4$, with $A_0$ such that $\max{(P)}=100\max{(b^2/2)}$. As we will see, this is sufficient to lead to the accumulation 
of a large magnetic flux on the black hole horizon, and a MAD accretion flow with $\beta\sim 1$.
We also add a random perturbation
to the internal energy density, $u = u_{\rm eq} + \delta u$, with $u_{\rm eq}$ the energy density in the Fishbone-Montcrief solution, and $\delta u$ drawn from
a uniform distribution covering $[-0.04 u_{\rm eq}, 0.04 u_{\rm eq}]$. This perturbation will seed turbulent motion in the disk. The pressure anisotropy and heat flux are
initialized to $\Delta P = q = 0$. The initial density and poloidal magnetic field of the SANE disk are shown in Fig.~\ref{fig:ID}.

\subsection{Numerical methods}
\label{sec:numerical-methods}

To evolve the EMHD model we developed a new code, \grim\footnote{General Relativistic \emph{Implicit} Magnetohydrodynamics: \url{http://github.com/afd-illinois/grim}}, described in~\cite{Chandra:2017}. Evolving the EMHD model
requires the use of implicit-explicit time stepping methods, instead of the purely explicit methods used to evolve the general relativistic
equations of ideal MHD. 
Indeed, the dissipative quantities in the EMHD model, namely the heat flux $q$ and the pressure anisotropy $\Delta P$, are sourced by spatio-temporal 
gradients of the thermodynamic variables.
In \grim, all flux terms in the equations are treated explicitly, while some local source terms are treated implicitly. Accordingly,
neighboring points are only coupled through terms treated explicitly, and the implicit solve can be done grid point by grid point. At each point and for
each time step, we have to solve a set of non-linear implicit equations for 7 evolved variables. The evolution of the magnetic field can be treated explicitly, and
fully decouples from the implicit solve. Numerical fluxes on cell
faces are computed with left- and right-biases stencils using the PPM reconstruction~\citep{PPM}, 
with a Local Lax-Friedrich approximate Riemann solver to reconstruct the fluxes from the left and right states on each face. 
Due to the use of implicit-explicit time stepping, the EMHD model is significantly more expensive to evolve than
IMHD. The simulations presented here can perform $\sim 10^5$
zone-cycles per second and per GPU node on the TACC Stampede cluster,
about two orders of magnitude slower than state-of-the-art explicit
evolutions on the same infrastructure (Liska, Tchekhovskoy, et al., in
preparation).

When evolving the equations of IMHD or EMHD in regions with relativistic velocities or with large magnetic fields, small discretization errors can
lead to large numerical errors in energetically subdominant physical variables (e.g. the velocity in magnetically dominated regions). To avoid this issue,
we implement techniques commonly used in IMHD simulations, imposing an upper bound on the Lorentz factor $\gamma$, 
and lower bounds on the density and internal
energy of the fluid. In particular, we require $\gamma < 10$, $\rho > 10^{-3} (r/r_{\rm in})^{(-5/2)}$, $u>10^{-5} (r/r_{\rm in})^{(-5/2)}$, $\rho > 0.1b^2$, 
and $u>0.002 b^2$. 
When these conditions are violated, we reduce the velocity and
increase $\rho$ or $u$ as needed. The addition of mass or energy is performed in the
drift frame of the plasma, as described in~\cite{Ressler:2016}.
At the inner boundary, 
these conditions are identical to those used in our previous axisymmetric simulations~\citep{Foucart:2016}. Farther away, the density and internal energy
floors decrease faster with radius than in previous simulations, as we found that in long 3D simulations higher floors could impact the evolution of low-density
outflows produced by the disk. 

We impose reflecting boundary conditions at the poles, an outflow boundary condition at the outer radial boundary ($u^r > 0$),
and periodic boundary conditions in the azimuthal direction. To avoid instabilities at the poles we also correct the value of the primitive variables in the two cells
closest to the polar axis. In these cells, the variables $(u^\theta, b^\theta, \Delta P, q)$ are assumed to go to zero linearly with the polar angle $\theta$, while the rest
of the hydrodynamic variables are assumed to be constant. We do not
modify the magnetic fields.

\subsection{Coordinates and grid structure}
\label{sec:coords}

The evolution equations for the EMHD model are solved on a 3-dimensional domain in modified Kerr-Schild coordinates $(t,a,b,c)$, 
related to the usual Kerr-Schild coordinates
$(t,r,\theta,\phi)$ by the maps
\beqn
r &=& \exp{a} \label{eq:r}\\
\tilde \theta &=& \arccos{\big([1-f(r)] \mu_{\rm Ref} + f(r) \mu_{\rm uni}\big)}\\
\mu_{\rm ref} &=& \cos{[\pi b + 0.35 \sin{(2\pi b)}]} \\
\mu_{\rm uni} &=& \cos{[\pi b]} \\
\phi &=& 2\pi c\\
f(r) &=& \frac{r_{\rm in}}{r} \label{eq:fr}
\eeqn
and a map $\tilde \theta \rightarrow \theta$ which derefines the grid in the $\theta$ direction close to the polar boundary. The exact form
of this map is provided in the appendix. The resulting grid structure is shown in Fig.~\ref{fig:ID}.

The mapping $\mu_{\rm ref}$ is taken from~\cite{Gammie2003}, and aims at focusing resolution close to the equatorial plane of the disk, where the MRI is more difficult
to resolve. We switch to a uniform mapping $\mu_{\rm uni}$ close to the inner boundary in order to avoid limiting the time step by the use of a very small
grid spacing in the $\theta$ direction on the boundary. The mapping from $\tilde \theta$ to $\theta$ serves a similar purpose: by derefining the grid 
close to the polar boundary, we avoid very small grid spacings in the azimuthal direction close to the axis. The map parameters are chosen
in order to maximize the minimum grid spacing across the entire domain.

\subsection{Resolution and convergence}

We evolve the SANE torus at two resolutions for both the IMHD and EMHD model. The high resolution simulations use $256\times 256 \times 128$ 
cells along the $(a,b,c)$ directions. The low resolution simulations use $128\times 128 \times 64$ cells. 
For the MAD disk, we use $256\times 128\times 64$ grid points
and evolve the system with both the EMHD and IMHD models.
In all simulations, we place the inner edge of the grid at $r_{\rm in}=0.85 r_{\rm H}$, with $r_{\rm H}=r_g (1+\sqrt{1-a^2})$ the radius of the black hole 
horizon. This guarantees 6 grid cells inside the horizon at low resolution.
The outer edge of the grid is at $r_{\rm out}=55 r_g$ for the SANE disk, and $r_{\rm out}=3000r_g$ for the MAD disk.
Finally, all SANE simulations are evolved for a time $\Delta t=6000 t_g$, and all MAD simulations for $\Delta t=14000 t_g$.

Before further analysis of our results, it is useful to note which effects our simulations can resolve. For the SANE disk, we rely for this analysis on the simulations at 
two numerical resolutions performed here, and on pre-existing ideal MHD simulations. In particular, we can compare our results with the detailed
analysis of an ideal MHD accretion flow presented in~\cite{Shiokawa2012}. ~\cite{Shiokawa2012} showed that simulations with numerical resolution comparable to 
our highest resolution were well converged for some global quantities (magnetization of the disk, temperature, accretion rate), and captured the scale of the fastest
growing mode of the MRI. However, they only marginally resolved the azimuthal correlation length of the fluid variables (density, internal energy), 
and underresolved the azimuthal correlation length of the magnetic field.
We find that our simulations have very similar properties. In particular, the differences observed in global, time-averaged disk quantities between our two resolutions
are comparable to the results at similar resolution in~\cite{Shiokawa2012}.
We also find that, at our highest resolution, the azimuthal correlation length of the fluid quantities is $\sim 7$ grid spacings, while the azimuthal correlation length of the
magnetic field is only $\sim 4$ grid spacings. We do not observe significant differences between the convergence properties of the EMHD and IMHD models.

From these results, we infer that our simulations allow us to do a direct comparison of the EMHD and IMHD models for a steady-state, SANE accretion flow.
We can compare time-averaged profile of the accretion flow, and the dynamics of the disk, but will not
fully capture the azimuthal variations in the disk or the turbulent cascade driven by the MRI. In that respect, our simulations are largely similar to existing IMHD results at
similar resolution. 

Because of the stronger magnetic fields in the MAD disks, their resolution
requirements are less stringent than for SANE disks. In particular,
with rather low resolutions, it is possible to resolve
the wavelength of the fastest growing mode of the MRI in the MAD disk
proper. Even though higher values of the azimuthal resolution,
$N_\phi$, lead to the emergence of smaller-scale structure in the
MAD disks, their steady state properties  appear to
be rather insensitive to the resolution, with $N_\phi = 64$
and $128$ giving qualitatively similar time-average disk
properties \citep{2011MNRAS.418L..79T}. However, higher
$\theta$-resolutions might be beneficial for
resolving the innermost regions of the MAD disks, in the immediate
vicinity of the event horizon, where the disks are strongly vertically
compressed by the magnetic pressure of the polar jets.

\section{Standard Disk Results}
\label{sec:SANE}

\subsection{Overview of the disk evolution}

The qualitative evolution of the standard (SANE) accretion flow studied here is largely independent of the underlying fluid model (IMHD vs EMHD), 
or of the numerical resolution used in the simulation.
The initial torus is in hydrostatic equilibrium, but unstable to the growth of the magnetorotational instability (MRI). Accordingly, the MRI grows from the seed perturbations
in our initial conditions, on a timescale comparable with the orbital timescale of the disk ($\sim 250 t_g$ at the pressure maximum of the initial torus).
\footnote{In linear theory, anisotropic pressure increases the growth rate of the MRI in the presence of a toroidal magnetic field, but not for a
purely poloidal magnetic field~\citep{Quataert2002,Balbus2004}.} 
The MRI drives
turbulence in the disk, causing outward angular momentum transport, the accretion of matter onto the black hole, and heating of the accretion flow. The initial
accretion torus then acts as a large matter reservoir feeding a quasi-steady state accretion flow onto the black hole. 
Steady-state is typically reached on a timescale
 $\tau_{st}(r)\sim \tau_d \alpha^{-1} \left(r/H\right)^2$, the ``viscous'' timescale of the disk, with $\alpha \sim 0.01-0.1$
and $H$ the scale height of the disk. 
In steady-state, a thick, 
turbulent accretion disk fills the equatorial region, while the polar
regions are magnetically dominated.

The initial transient leading to the formation of a steady-state accretion flow is mostly determined by our choice of initial conditions, 
and the exact realization of the seed perturbation. 
We will largely ignore that phase in our analysis. The time-averaged properties of the steady-state accretion flow, on the other hand, are more robust, and thus more
interesting to study. This does not mean, however, that our chosen initial conditions no longer have an impact. In particular, there are two important aspects of our initial
conditions which should be kept in mind when analyzing our results. 

The first is that the initial equilibrium torus for the SANE
simulation is quite compact, and resides deeply in the gravitational potential of
the black hole. This means that the accretion flow at distances $r \gtrsim 10 r_g$ remains, at all times, solely a consequence of the initial conditions. 
Additionally, disk outflows are artificially suppressed in this configuration: all of the matter starts with a relatively high binding 
 energy, and close to the black hole. The second important impact of the initial conditions is the evolution of the net magnetic flux threading the black hole horizon, which
 determines whether the disk ends up in the SANE or MAD configuration. We discuss an accretion flow reaching the MAD state in Sec.~\ref{sec:MAD}.

\begin{figure}
\includegraphics[width=\columnwidth]{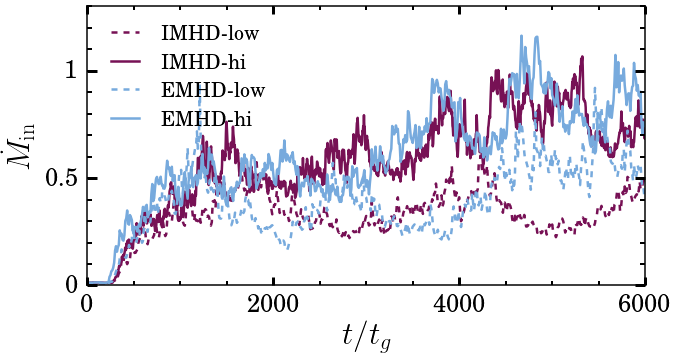}
\caption{Accretion rate across the black hole horizon as a function of time, for the EMHD and IMHD SANE models at low and high resolution.
Note the very similar accretion history for the two high-resolution runs. }
\label{fig:MdotTime}
\end{figure}

The rate $\dot{M}$ of accretion of matter onto the black hole for each of our 4 SANE simulations is shown in Fig.~\ref{fig:MdotTime}. In all cases, accretion begins after
a time delay consistent with the MRI growth timescale, then reaches a
turbulent quasi-steady state on a timescale of the order of the
viscous timescale at the innermost stable circular orbit of the black
hole. In the rest of this paper, we focus our SANE disk simulation analysis on the time period $(4000-6000)t_g$, when we expect quasi-steady accretion
for $r \lesssim 7r_g$. We see that during that period, the average accretion rates of the high-resolution IMHD and EMHD simulations are very similar, which will facilitate
comparisons between the two models. The low-resolution simulations have more significant differences in their accretion history, but these remain consistent with random
variations of the accretion rate over the relatively small timescales that we can afford to simulate.

\subsection{Steady-state disk profile}

\begin{figure}
\includegraphics[width=\columnwidth]{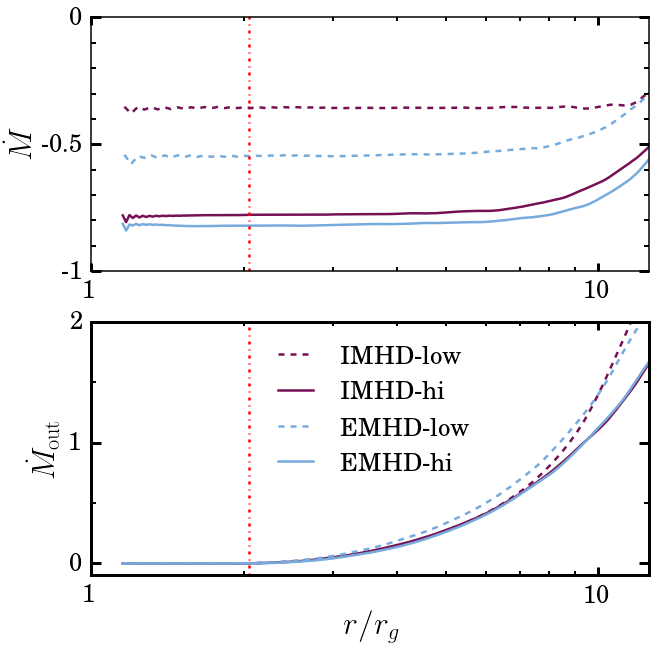}
\caption{Time averaged mass accretion rate $\dot{M}$ as a function of radius for the SANE accretion flow.
The vertical dot-dashed red-lines show the location
of the innermost stable circular orbit (ISCO) for point-particles in prograde orbits.
{\it Top}: Net accretion for the IMHD and EMHD models at two resolutions.
We see that all disks reach a quasi steady-state up to $r\sim
(8-9) r_g$. 
{\it Bottom}: Total outgoing mass flux for the IMHD and EMHD models at two
resolutions (i.e. integrating the radial mass flux solely over regions
in which it is positive). 
This mass flux is due to turbulent, outwardly directed motion of gas in the disk.
  }
\label{fig:MdotRad}
\end{figure}

\begin{figure}
\includegraphics[width=0.9\columnwidth]{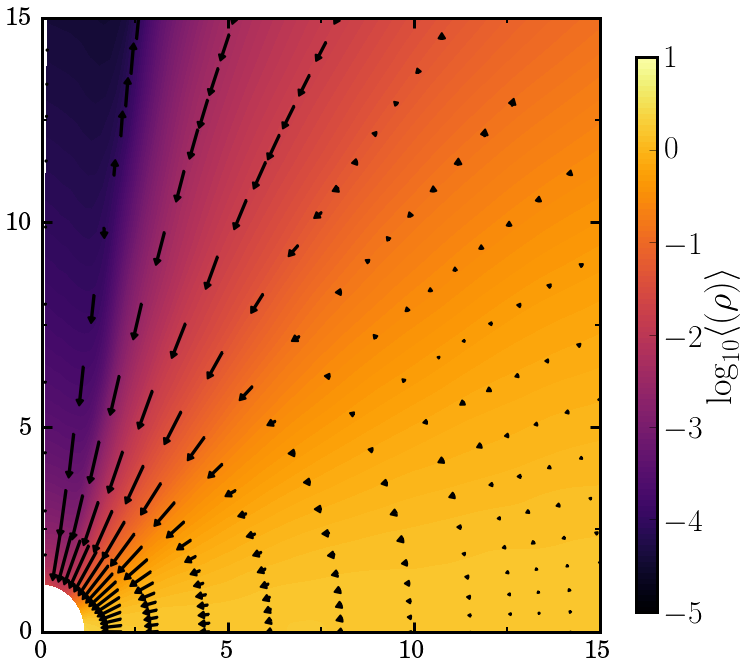}
\caption{Density and velocity of the SANE accretion flow in the EMHD model, averaged over the azimuthal direction and within the time window
$t=(4000-6000)t_g$. There are no steady-state outflows in this model (the outflows at $r>10r_g$ are not in the steady-state region).
All distances are in units of $r_g = GM/c^2$.}
\label{fig:velSANE}
\end{figure}

Fig.~\ref{fig:MdotRad} shows the net accretion rate $\dot{M}$, and the outgoing mass flux $\dot{M}_{\rm out}$ as a function of radius, time-averaged over the window
$t=(4000-6000)t_g$. $\dot{M}_{\rm out}$ is computed by only integrating the radial mass flux over fluid elements with outgoing radial velocities. The constant
$\dot{M}$ profiles for $r \lesssim 7 r_g$ confirm that we have reached a quasi-steady state in that region, and that we are in fact very close to steady-state up
to $r\sim 9 r_g$. Nearly all of the matter is flowing inwards for $r \lesssim (3-4) r_g$, while at larger radii turbulent motions in the disk cause some material
to move outwards, with $\dot{M}_{\rm out} \sim |\dot{M}|$ in the outermost regions of the steady-state accretion flow. As shown by the velocity streamlines
in Fig.~\ref{fig:velSANE}, we do not observe any spatial region with steady-state outflows, confirming the absence of any strong disk wind in the compact SANE disk considered here.

\begin{figure}
\includegraphics[width=\columnwidth]{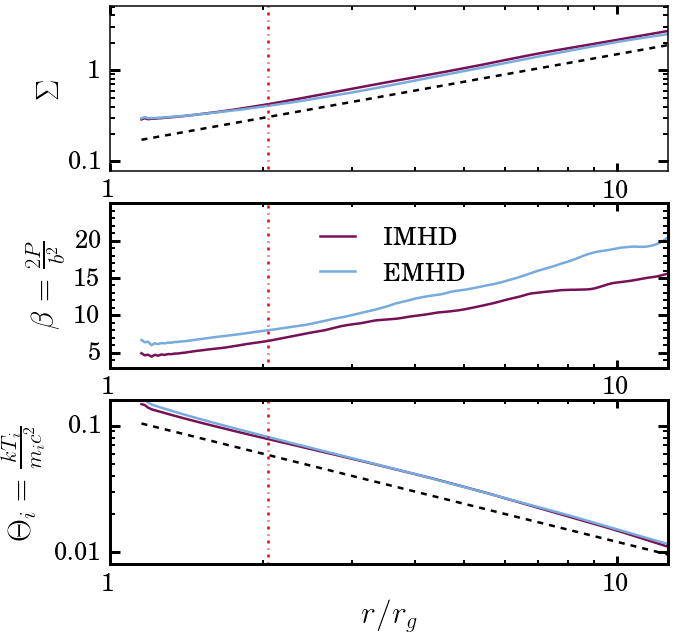}
\caption{Surface density $\Sigma$, plasma parameter $\beta=2P/b^2$, and dimensionless ion temperature $\Theta_i$ 
as a function of radius for the SANE configuration  using
the EMHD and IMHD models. 
$\Sigma$ is averaged over time and the azimuthal direction. For $\beta$, we compute the ratio of the density-weighted averages of the gas pressure
and magnetic pressure, while we directly compute the density-weighted average value of $\Theta$. 
$\beta$ is higher in the EMHD simulations, while the gas temperature and surface density are nearly identical in both models.
The dashed black lines show power laws $\Sigma \propto r$ and $\Theta_i \propto r^{-1}$, while the dot-dashed red-lines show the location
of the ISCO.}
\label{fig:FluidProf}
\end{figure}

Some of the most important properties of the accretion flow in steady-state are shown in Fig.~\ref{fig:FluidProf}, which shows radial profiles of the surface density
$\Sigma$, ion temperature $\Theta_i$, and plasma parameter $\beta = 2 P/b^2$ (the ratio of the gas to magnetic pressure). These profiles are averaged over the time
window $t=(4000-6000)t_g$, and over the azimuthal angle. They are also integrated over the polar angle $\theta$. Unless otherwise specified, all figures show results
for the higher resolution simulations.
We see that the surface density and ion temperature in the disk are nearly identical for the IMHD and EMHD models. They also
closely follow the power-law profiles $\Sigma \propto r$ and $\Theta_i \propto r^{-1}$. The latter is consistent with the conversion of gravitational 
potential energy into thermal energy, as expected for a radiatively inefficient disk. On the other hand, when it comes to the plasma $\beta$, the IMHD and EMHD 
model differ: the disk is more strongly magnetized in ideal MHD, with $\beta_{\rm EMHD} \approx (1.2-1.4) \beta_{\rm IMHD}$. 

Overall, the most striking feature of the steady-state disk profiles are the similarities between the IMHD and EMHD simulations, shown for the fluid variables ($\rho$, $\Theta_i$) in the radial profiles shown in Fig.~\ref{fig:FluidProf}, but also verified
for angular profiles at fixed radius, or for the azimuthal correlation length of the fluid variables and
of the magnetic field. Despite the significant pressure anisotropy in the disk (discussed below), the steady-state accretion flow is remarkably independent of the choice
of fluid model. The only difference which cannot be explained by
random fluctuations in the accretion flow is the lower magnetic field
strength in the EMHD simulation.

\subsection{Pressure anisotropy and heat conduction}

\begin{figure}
\includegraphics[width=\columnwidth]{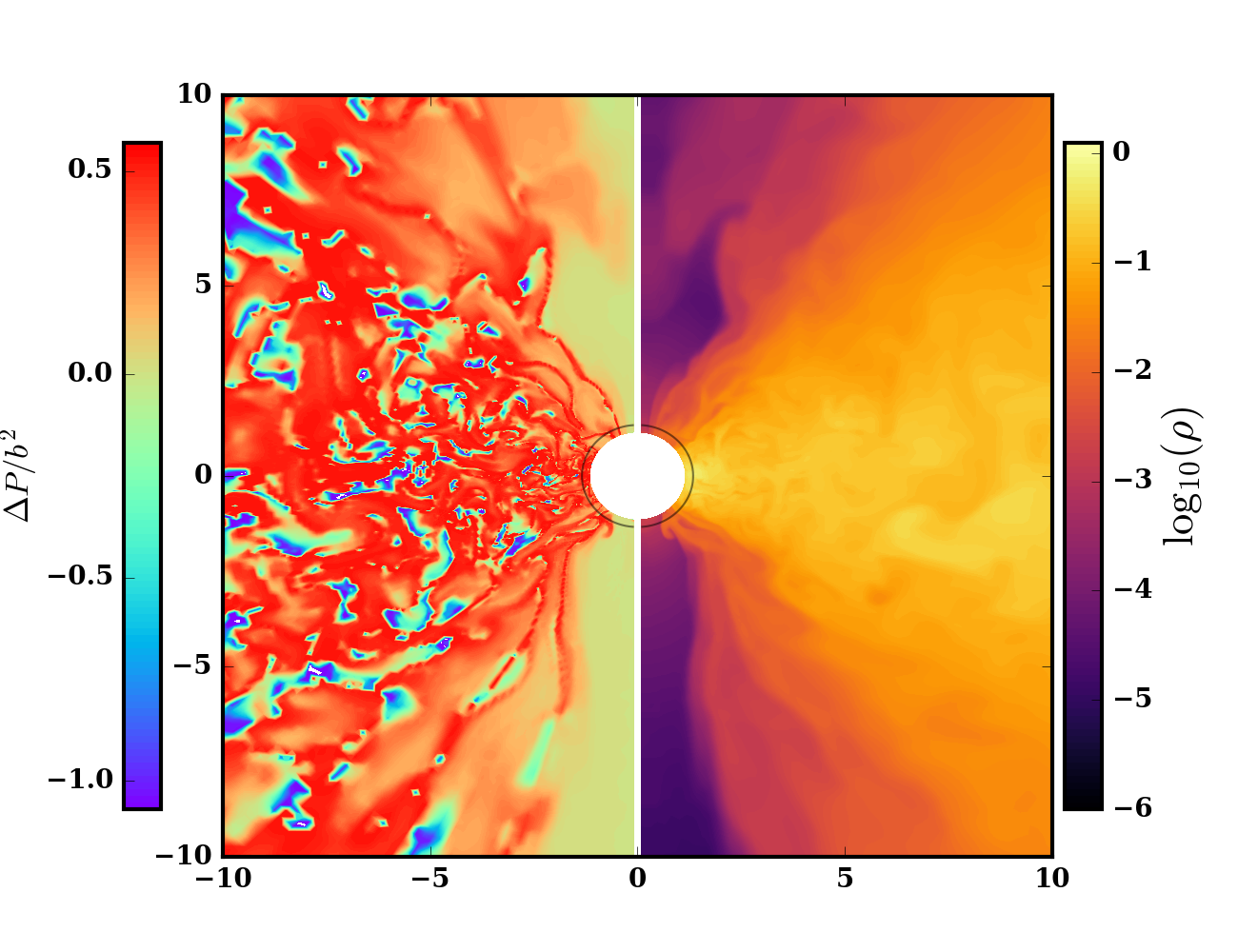}
\caption{Vertical slice through the high-resolution EMHD model at $t=5000 t_g$ (SANE disk). {\it Left}: Pressure anisotropy 
normalized by the magnetic energy density. The pressure anisotropy saturates at the threshold of the mirror instability, 
$\Delta P \sim b^2/2$, in most of the disk.
{\it Right}: Fluid density, showing the magnetically driven turbulence.
}
\label{fig:EMHDSlice}
\end{figure}

In order to understand the impact of non-ideal effects on the dynamics of the disk, it is useful to look at the instantaneous and average value of the pressure
anisotropy $\Delta P$. In Sec.~\ref{sec:emhd}, we noted that for a Keplerian velocity profile we expect the pressure anisotropy to grow to
the threshold of the mirror instability, $\Delta P \approx b^2/2$, as long as $\beta > 1$. In practice, our EMHD simulations show that this is the case
even in the turbulent accretion flow resulting from the MRI-driven evolution of the disk. 

To illustrate this, we show in Fig.~\ref{fig:EMHDSlice} a specific
realization of the pressure anisotropy $\Delta P$ for a vertical slice
of the SANE disk at $t=5000 t_g$.
In most of the accretion disk (i.e. away from the magnetically dominated polar regions, where $\beta \ll 1$), we find that $\Delta P \approx b^2/2$, the threshold of the
mirror instability. Some small portions of the disk instead find themselves at the threshold of the firehose instability, $\Delta P \approx -b^2$.

\begin{figure}
\includegraphics[width=\columnwidth]{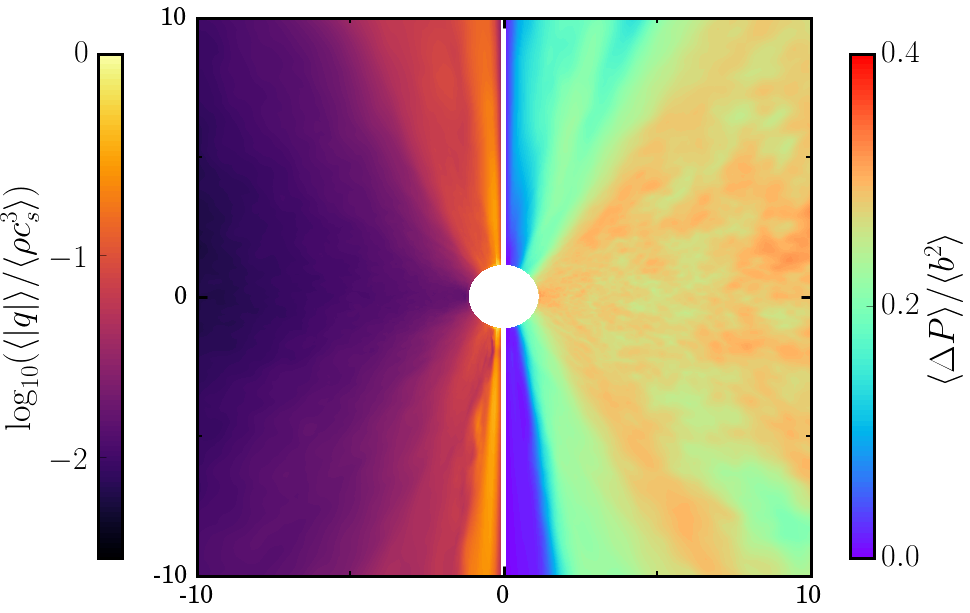}
\caption{Time and azimuthally-averaged profiles for the high-resolution EMHD model (SANE disk). 
{\it Left}: Ratio of the averages of $|q|=\sqrt{q^\mu q_\mu}$
and $q_{\rm max}=\rho c_s^3$. {\it Right}: Ratio of the averages of $\Delta P$ and $b^2$. While the heat flux is far
from saturated in the disk (typically a few percent of its saturation value), the average pressure anisotropy is of order
the magnetic pressure, $\sim (0.2-0.4) b^2$.
}
\label{fig:EMHDSlice2}
\end{figure}

\begin{figure}
\includegraphics[width=\columnwidth]{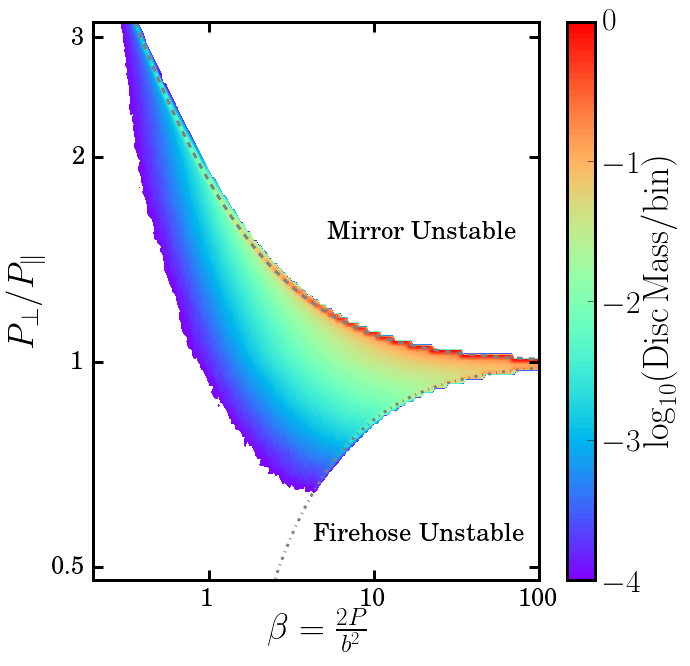}
\caption{Mass-weighted distribution function in the $\beta$-$(P_\perp/P_\parallel)$ plane (SANE disk).
The dashed curve shows the threshold of the mirror instability,
and the dot-dashed curve the threshold of the firehose instability. Most of the mass remains at the mirror-instability threshold. We show the total mass
integrated over 100 equidistant snapshots of the simulation between $(4000-6000)GM/c^3$. The bins are logarithmically spaced in both dimensions,
and we only include points at $r<9 GM/c^2$, where the flow is in quasi-steady state.
}
\label{fig:dPdis}
\end{figure}

\begin{figure}
\includegraphics[width=\columnwidth]{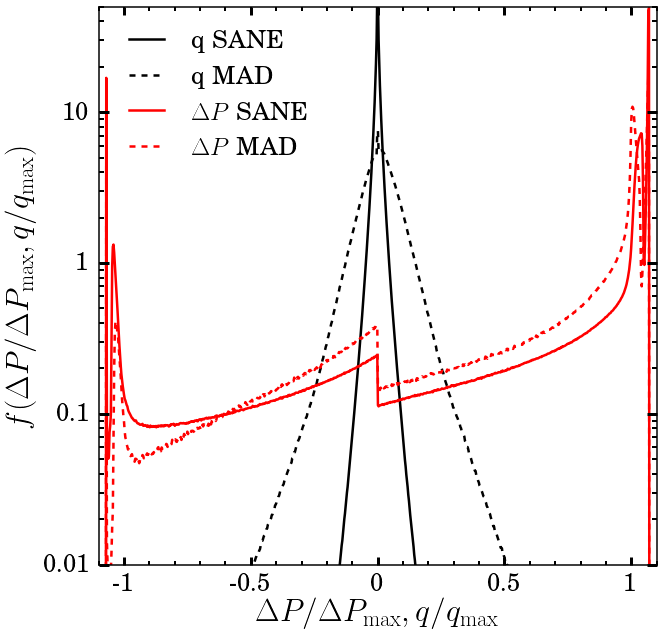}
\caption{Mass-weighted distribution function of the normalized pressure anisotropy ($\Delta P/\Delta P_{\rm max}$, with $\Delta P_{\rm max}$ the threshold
of the mirror/firehose instability for positive/negative $\Delta P$) and normalized heat flux ($q/q_{\rm max}$, with $q_{\rm max}=\rho c_s^3$) 
for the SANE (solid lines) and MAD (dashed lines) configurations. 
Most of the fluid shows a large pressure anisotropy $\Delta P\sim \Delta P_{\rm mirror}$, and a small heat flux. The discontinuity in the distribution of $\Delta P$ at
$\Delta P=0$ is due to the different normalization for positive/negative $\Delta P$.
}
\label{fig:1Ddis}
\end{figure}

The average pressure anisotropy in the disk is shown in Fig.~\ref{fig:EMHDSlice2}. Once we average $\Delta P$ in time and over
the azimuthal direction, we find that $\langle \Delta P \rangle \approx (0.2-0.4) \langle b^2 \rangle$.
This is consistent with the more detailed distributions of $\Delta P$ provided in Figs.~\ref{fig:dPdis}-\ref{fig:1Ddis}: in steady-state, about $65\%$ of the
disk mass is close to the mirror instability threshold, at $\Delta P \gtrsim 0.99 \Delta P_{\rm mirror}$. About $5\%$ of the disk mass is
close to the firehose instability threshold, and the average value of $\Delta P/b^2$ is $0.25$. In earlier axisymmetric simulations, we already
noted that about half of the disk was at the threshold of the mirror instability~\citep{Foucart:2016}. These results show that the fluid is pushed 
even more towards the mirror instability threshold in 3D simulations.

In contrast to these large pressure anisotropies, 
heat conduction in the disk remains negligible throughout the simulation. Fig.~\ref{fig:EMHDSlice2} and Fig.~\ref{fig:1Ddis} 
show that the magnitude of the heat flux in
the disk is, on average, only a few percent of the free-streaming value $q_{\rm max}=\rho c_s^3$. 
The heat flux only becomes larger in the polar regions, where magnetic fields
dominate the evolution of the system. 
This is consistent with results obtained in axisymmetry, where the low value of the heat flux was due to (i) the saturation
of the pressure anisotropy, which causes a suppression by a factor of $\beta^{-1}$ of the collision timescale $\tau_R$, and thus of the heat flux; and (ii) the misalignment
between the poloidal temperature gradient and the largely azimuthal magnetic field. As heat conduction can only act along magnetic field lines, temperature gradients
orthogonal to the magnetic field cannot drive heat conduction in the disk. Fig.~\ref{fig:EMHDSlice} shows that, even in 3D, the heat flux is 
negligible in regions in which $\Delta P$ saturates.

From our EMHD simulation, we thus deduce that the average impact of the non-ideal effects in the compact, SANE disk studied here
might be close to what one would obtain by setting $\Delta P \sim 0.25 b^2$ and $q\sim 0$. We also note that our results critically depend
on the treatment of the mirror instability in the EMHD model. We assume that when driven above the threshold of the mirror instability, $\Delta P$ saturates
at nearly exactly the instability threshold $\Delta P_{\rm mirror}$, through an increase in the effective collision rate in the plasma. Our results are only
physical if this approximation captures the global behavior of the plasma on length scales much larger than the Larmor radius. 
While current particle-in-cell simulations broadly support this model~\citep{Kunz:2016}, comparisons of the EMHD
model with more detailed local simulations of a plasma at the mirror
instability threshold will be required to test this assumption. 

\subsection{Stress tensor and angular momentum transport}

In previous sections, we showed that the pressure anisotropy in the EMHD simulations of SANE disks is, in a time average sense, $\Delta P \sim b^2/4$.
In Sec.~\ref{sec:emhd}, we also argued that the associated shear stress, $\sim -\Delta P \hat{b}^\mu \hat{b}^\nu$, is effectively of the same
form as the magnetic shear stress, $-b^2 \hat{b}^\mu \hat{b}^\nu$. As far as the shear stress is concerned, the pressure anisotropy $\Delta P$ and the magnetic energy
density $b^2$ are interchangeable. Additionally, for $\beta \gg 1$, the magnetic energy density and isotropic magnetic pressure do not play a particularly important
role in the dynamics of the accretion flow.

\begin{figure}
\includegraphics[width=\columnwidth]{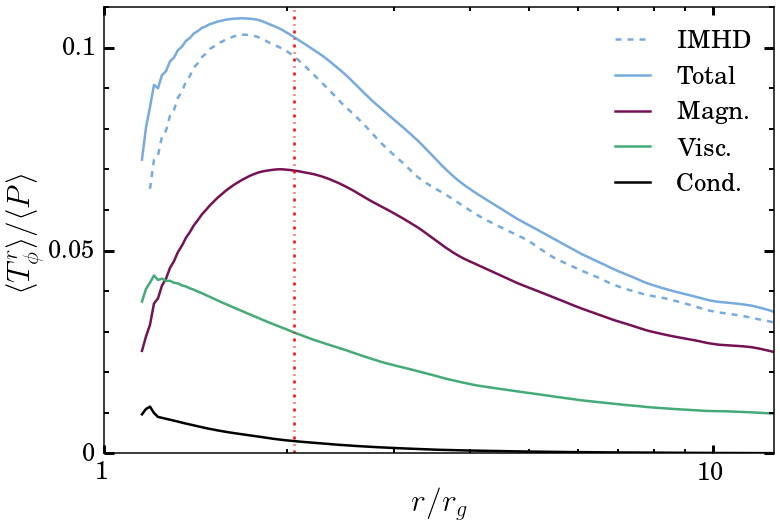}
\caption{Average (r,$\phi$) component of the stress tensor, $\langle T^r_\phi \rangle$, normalized by the average pressure $\langle P \rangle$ in the SANE simulations. 
The dashed curve shows the magnetic stress in the IMHD simulation. The EMHD stress is decomposed into magnetic, viscous, and 
conduction components. The total stress is nearly the same in the EMHD and IMHD simulations. 
The lower value of the magnetic stress in the EMHD model is compensated by the viscous stress. The red dot-dashed line shows the location of the ISCO.
The turnover of the magnetic stresses
at the ISCO is due to the growing contribution of the advection term ($\sim b^2 u^r u_\phi$), which is negligible at large radii, does not have a viscous
counterpart, and is transporting angular momentum inwards.}
\label{fig:stress}
\end{figure}

Accordingly, we may expect similar steady-state accretion flows to be reached in EMHD and IMHD if $b^2_{\rm IMHD}=b^2_{\rm EMHD} + \Delta P_{\rm EMHD}
\approx (5/4) b^2_{\rm EMHD}$.
Our simulations appear consistent with this interpretation of the impact of the pressure anisotropy: in the SANE configuration, we find similar density
and temperature profiles for the EMHD and IMHD models, but with a weaker magnetic field in the EMHD model, e.g.
the plasma parameter $\beta$ satisfies $\beta_{\rm EMHD} = (1.2-1.4)\beta_{\rm IMHD}$. 

We can have a more direct look at the component of the stress-energy tensor responsible for angular momentum transport in the disk, 
$T^r_\phi$. We show radial profiles of
$T^r_\phi$ (averaged over time and azimuthal angle, integrated over the polar angle, and normalized by the gas pressure) in Fig.~\ref{fig:stress}. 
The first notable result is that the total stress is nearly identical
for the IMHD and EMHD simulations. However, in the IMHD model the entirety of the stress is due to the magnetic field. In the EMHD model, outside the innermost stable
circular orbit of the black hole, $(70-75)\%$ of the $r-\phi$ component of the stress is due to the magnetic field and $(25-30)\%$ to the pressure anisotropy. While this
fraction of the stress due to the pressure anisotropy is slightly larger than one might have expected based on the simplistic model of a constant ratio $\Delta P:b^2=1:4$,
the properties of the stress tensor are broadly consistent with the view that in EMHD the pressure anisotropy effectively replaces a fraction of the IMHD
magnetic field, with no obvious other significant effects on the dynamics of the disk.

Finally, we note that the profiles for the effective $\alpha$-viscosity (defined by $\langle T^r_\phi \rangle = \alpha \langle P\rangle$), surface density 
$\Sigma$, and temperature $T$ shown in Fig.~\ref{fig:FluidProf} and Fig.\ref{fig:stress} do not agree with the usual 1D result $\alpha \Sigma \Theta/\Omega = {\rm constant}$.
This can be understood from the conservation of angular momentum equation (expressed, for simplicity, in the Newtonian limit and for a quasi-Keplerian
thin disk),
\beq
\dot{M} r^2 \Omega + 2\pi \alpha r^2 \Sigma \Theta = \dot{J} \label{eq:Jcons},
\eeq
with $\dot{J}$ the net angular momentum transported through the disk, which is constant in the steady state region. As an order of magnitude
estimate, we should have $\dot{J} \sim \dot{M} j_{\rm ISCO}$, with $j_{\rm ISCO}$ the specific angular momentum at the ISCO,
and thus $\dot{M} j(r) \sim (1-2) \dot{J}$ within the steady-state accretion flow in our simulations. The relation $\alpha \Sigma \Theta/\Omega = {\rm constant}$
is only valid when $\dot{M} j(r) \gg \dot{J}$, which is typically the case at larger radii but not close to the ISCO. At small radii and outside the ISCO, we instead expect
$\Sigma$ to grow as $(r-r_{\rm ISCO})$ increases, as seen in our simulations. 

\section{MAD Results}
\label{sec:MAD}

The magnetically arrested disk (MAD) simulations provide us with a different regime to study the impact of non-ideal effects. 
The MAD simulations are initialized using a wider torus in which we seed a single loop of poloidal magnetic field
(see Sec.~\ref{sec:ID}). The vertical magnetic flux crossing the equatorial plane
inside of the pressure maximum of the disk is significantly larger in the MAD configuration than in the SANE configuration.
This vertical flux is slowly accreted onto the black hole, up to the point at which
the magnetic flux threading the black hole becomes large enough to
vertically compress the disk and obstruct the gas infall. This happens when the magnetic 
pressure from the jet is in equilibrium with the ram pressure of the infalling gas. 
The accretion flow close to the horizon then proceeds through a very thin
disk. That disk is more strongly magnetized than in the SANE case ($\beta \sim 1$ at the ISCO), and the magnetic pressure of the
jet plays an important role in its evolution. With a dynamically important magnetic pressure, we expect larger differences between the 
magnetic and viscous stresses than in the SANE disk. However, we will see that this does not translate into significant differences
between the EMHD and IMHD results. Accordingly, we will not repeat here the full analysis of the accretion flow provided for the 
SANE case, and will instead only  discuss the main differences between the MAD and SANE results. 

 In order to determine when we enter the MAD state, we monitor the normalized flux across the horizon,
 \beq
\phi_{\rm BH} = \sqrt{4\pi}\frac{\oint |B^a| \sqrt{-g} dS}{2\dot{M}^{1/2}},
\eeq
with the surface integral taken on the event horizon
of the black hole, $a=\log{(r)}$,
and $g$ the determinant of the metric $g_{\mu \nu}$ in the coordinates $(t,a,b,c)$ of the numerical grid (see Sec.~\ref{sec:coords}). 
$\phi_{\rm BH}$ grows steadily up to $t\sim (5-6) 10^3 t_g$, then oscillates around
the average value $\bar \phi_{\rm BH}=75$ in EMHD, and $\bar \phi_{\rm BH}=74$  in IMHD for times
$t>6000t_g$.~\footnote{
 $\bar \phi_{\rm BH}$ is computed by taking time-averages of $\oint |B^a| \sqrt{-g} dS$ and $\dot{M}$ separately.}
 This is comparable
to the value measured for the same black hole spin in~\cite{Tchekhovskoy2012} ($\bar \phi_{\rm BH}=60$). The difference
between our results and~\cite{Tchekhovskoy2012} is most likely due to the use of a different equation of state, resulting
in a thicker disk in this work. We note that the remarkable agreement in $\phi_{\rm BH}$ between the
IMHD and EMHD simulations occurs despite the accretion rate being $20\%$ lower for the EMHD model.

\begin{figure}
\includegraphics[width=0.9\columnwidth]{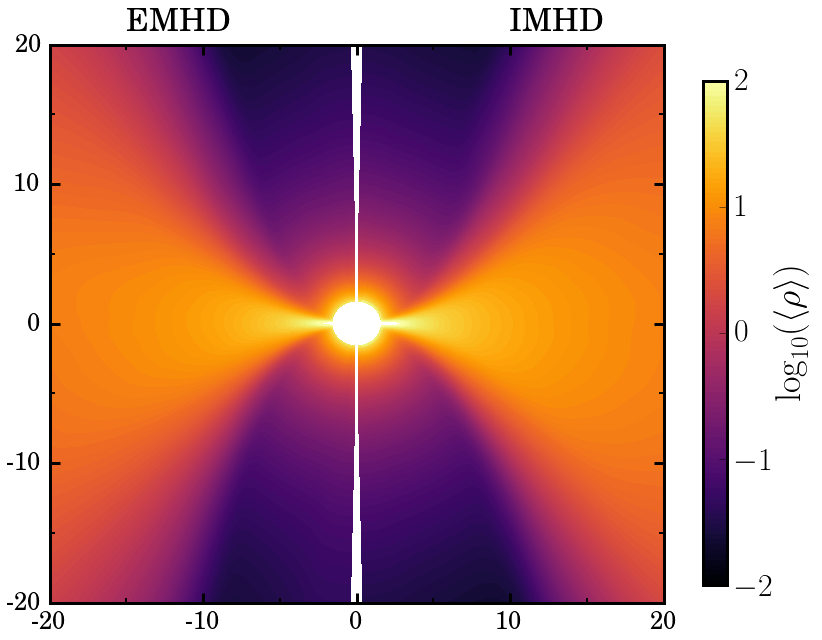}
\caption{Density of the MAD accretion flow in the EMHD (left) and IMHD (right) models, averaged over the azimuthal direction 
and within the time window $t=(6000-14000)t_g$. A magnetically arrested accretion flow is observed in both models.}
\label{fig:rhoMAD}
\end{figure}

\begin{figure}
\includegraphics[width=0.9\columnwidth]{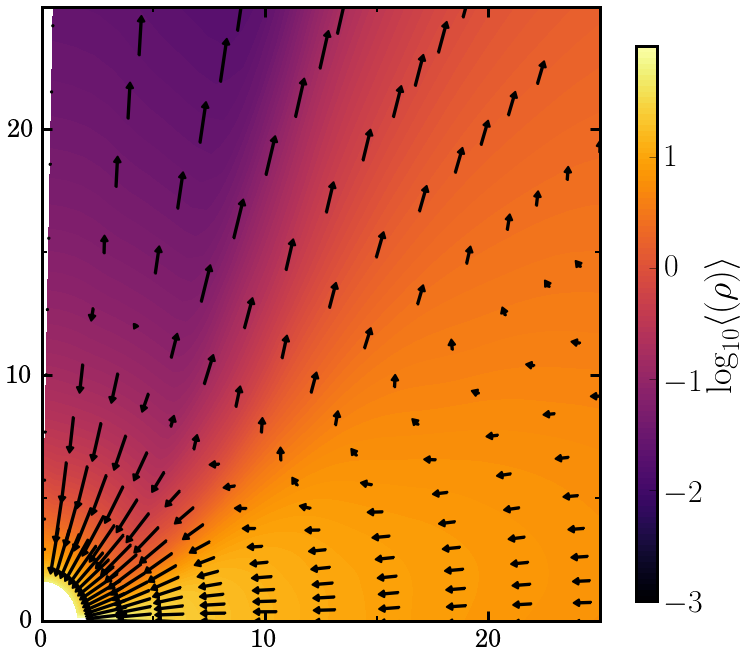}
\caption{Density and velocity of the MAD accretion flow in the EMHD model, averaged over the azimuthal direction and within the time window
$t=(6000-14000)t_g$. As opposed to the SANE model, we observe a clear inflow-outflow structure in the accretion flow.}
\label{fig:velMAD}
\end{figure}

\begin{figure}
\includegraphics[width=0.95\columnwidth]{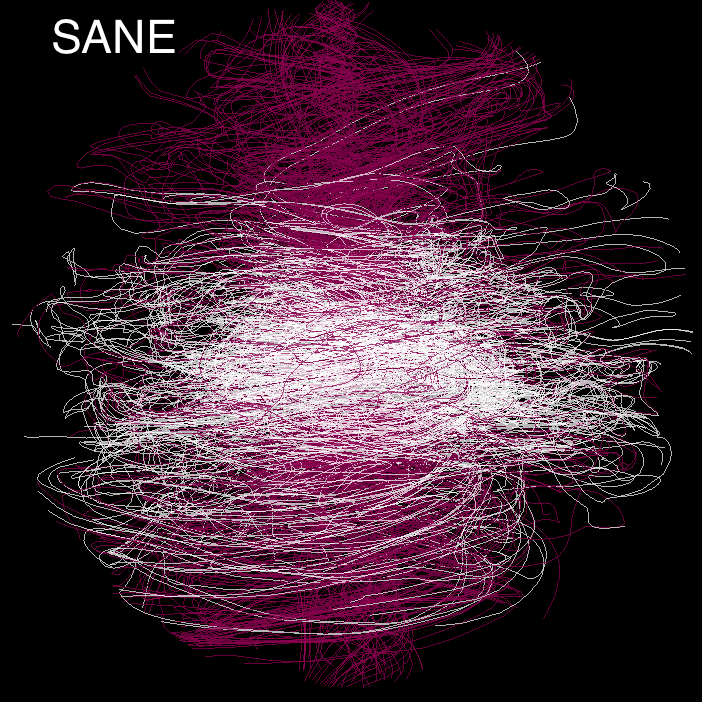}\\
\includegraphics[width=0.95\columnwidth]{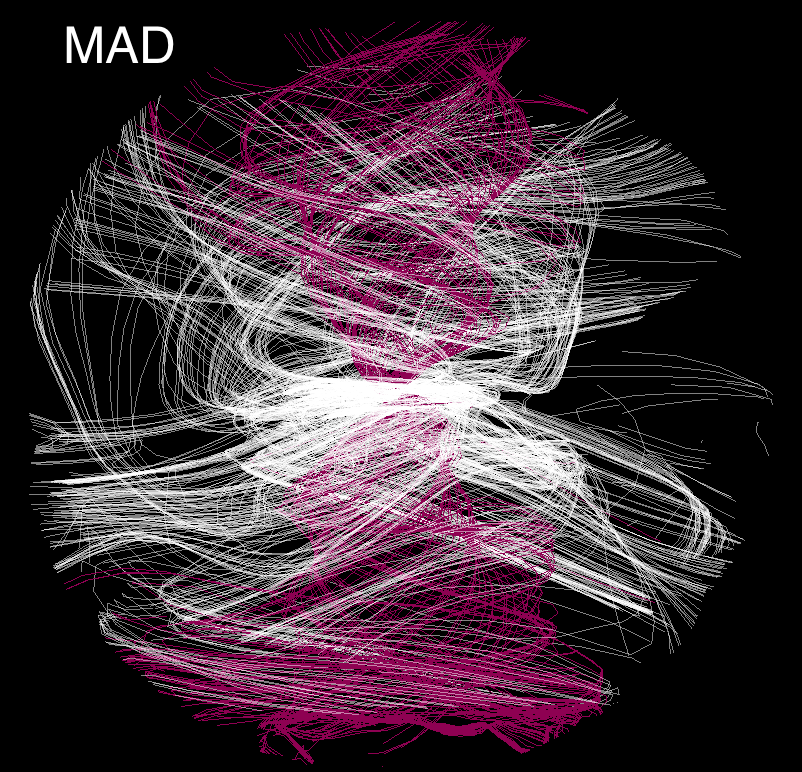}
\caption{Magnetic streamlines for the SANE ({\it top}) and MAD ({\it bottom}) simulations, using the EMHD model.
White streamlines are seeded in the equatorial plane, while purple streamlines are seeded above and below that plane
(at $z = \pm 10r_g$ for SANE, $z=\pm 20r_g$ for MAD). In all cases, the streamlines are randomly seeded within a
rectangular region centered on the polar axis in which the disk is in steady state [squares of size $(10r_g)^2$ for the SANE simulation and
$(20r_g)^2$ for the MAD simulation]. We stop integration of the streamlines at a distance of $50r_g$ from the black hole. 
The magnetic field is more structured in the MAD simulation, with clear wind and jet regions.}
\label{fig:Bfields}
\end{figure}

In the rest of this section, we focus on the time
interval $t=(6-14)10^3 t_g$ during which the simulations are in the MAD state, 
and compute time-averaged quantities over that interval.
Time-averaged density profiles of the steady-state accretion flow are shown in Fig.~\ref{fig:rhoMAD}
for the EMHD and IMHD models. Both simulations clearly show a
vertically compressed accretion flow, indicative of the MAD state \citep{2011MNRAS.418L..79T}, 
instead of the thick accretion flow observed in the SANE configuration. The EMHD and IMHD results are very similar.
While we did not find any steady-state disk outflows in the SANE configuration, a clear inflow-outflow structure
appears in the MAD simulations, as shown in
Fig.~\ref{fig:velMAD}. The inflowing fluid covers the region within $\sim 40^\circ$ of the
equatorial plane, while outflows are present at higher latitudes. The magnetically dominated jet itself covers the region
$10^\circ-20^\circ$ away from the poles. The net average mass flow
across spheres of fixed radii is constant up to $r\sim 50r_g$. The MAD disk has a larger steady-state region than the SANE disk
due to efficient angular momentum transport by large scale magnetic stresses (see Fig.~\ref{fig:Bfields}, and discussion below).

\begin{figure}
\includegraphics[width=\columnwidth]{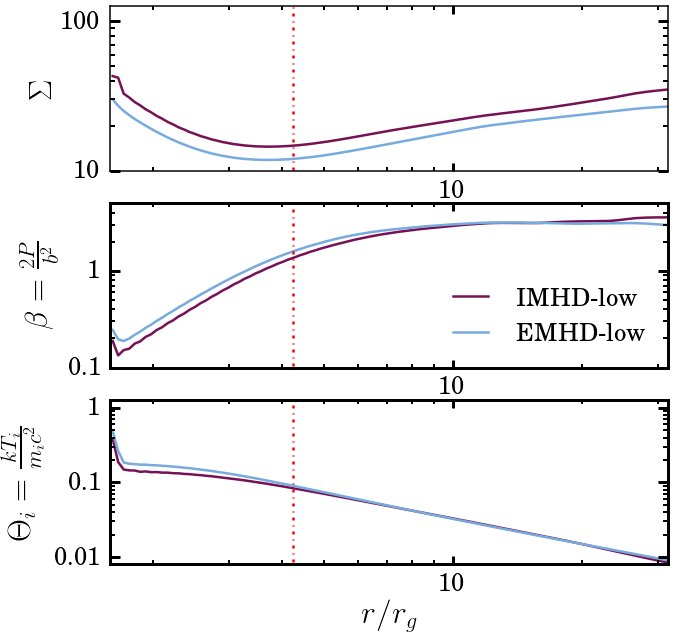}
\caption{Same as Fig.~\ref{fig:FluidProf}, but for the MAD configuration and with time averages taken over the $t=(6000-14000)t_g$ interval.}
\label{fig:FluidProfMAD}
\end{figure}

Fig.~\ref{fig:FluidProfMAD} shows one-dimensional profiles of the accretion flow for the surface density, plasma $\beta$-parameter, and fluid
temperature. In the SANE disks, the sole difference between EMHD and IMHD results was in the profile of $\beta$: the disk
was more strongly magnetized in the IMHD model than in EMHD. In the MAD disks, $\beta$ and $\Sigma$ both show deviations
of $15\%-25\%$, with the lower surface density being consistent with the $20\%$ slower accretion rate measured for the EMHD model. 
The temperature does not change by more than $10\%$ between the two models.

Finally, Fig.~\ref{fig:1Ddis} shows the pressure anisotropy and heat flux in the EMHD simulation. As in the SANE simulations, the pressure
anisotropy $\Delta P$ saturates at the threshold of the mirror instability. The heat flux is larger for MAD disks than SANE disks, of the order
of $10\%$ of its free-streaming value in large regions of the disk. This can be attributed to the higher magnetization of the MAD disk. As discussed
in Sec.~\ref{sec:emhd}, we expect an effective collision timescale $\tau_R = \tau_d/\beta$ in the disk, which causes a suppression of the heat
flux by a factor of $\beta$. While in the SANE disk we have $\beta\sim 5-15$, here we have $\beta\sim 1-3$ outside of the ISCO. The 
heat flux in the MAD disks is thus larger, yet remains far from the free-streaming value nearly everywhere. 

As in the SANE disk, the pressure anisotropy
contributes to outward angular momentum transport in the disk, with the vertically averaged shear stress satisfying 
$\langle T^r_{\phi,{\rm vis}}\rangle \sim 0.04 \langle P\rangle$. Near the equatorial plane, the magnetic shear stress
$T^r_{\phi,{\rm mag}}\sim (0.10-0.15) P$, so that the viscous shear stress is still $\sim 1/3$ of the magnetic shear stress, as
in the SANE simulation. However, in the MAD
configuration, there is a net extraction of angular momentum from the black hole, and most of that angular momentum is carried outward in the wind and, even 
more, in the jet.
In these magnetically dominated regions, $\Delta P \sim P \ll b^2$. Accordingly, we find that the contribution of magnetic fields to angular momentum
transport is $\sim 5-10$ times larger than the contribution of the pressure anisotropy in most of the steady-state region. 
In addition, the mass and angular momentum outflows, as well as
$T^r_{\phi,{\rm mag}}$, are similar in the EMHD and IMHD simulations.

Overall, we thus find that the EMHD and IMHD results are very similar for the MAD disk. While the viscous shear stress due to the pressure anisotropy
is larger for the MAD configuration than for the SANE configuration, the transport
of angular momentum in the jet and winds due to the magnetic field dominates over the viscous shear stress. 

\section{Conclusions}
\label{sec:conclusions}

We have presented the first global, 3D simulations including the two non-ideal effects most likely to impact
the dynamics of accretion disks around slowly accreting supermassive black holes: heat conduction along magnetic field lines, and an anisotropy
between the pressure parallel and perpendicular to the magnetic field direction. The latter is equivalent to an anisotropic viscosity.
We start from well-known configurations for an accretion torus in hydrostatic equilibrium,
which allows us to generate steady-state accretion flows in the region close to the black hole. We consider two initial configurations. The first
leads to a steady-state accretion driven by magnetically-driven turbulence and viscous stresses, with the turbulence driven by the growth of the 
magnetorotational instability. Angular momentum transport in the disk is due to both the magnetic  stress 
and the pressure anisotropy. The initial magnetic field has a relatively small net vertical magnetic flux, so that no significant magnetized jet forms along the black hole
rotation axis. The disk remains in the so-called SANE configuration, with a plasma parameter $\beta \sim 5-15$.
In the second configuration, we consider a wider initial torus with a larger net vertical magnetic flux. As the magnetic flux across the black hole horizon increases, 
the magnetic pressure of the jet balances the ram pressure of the accreting gas, forming a magnetically arrested disk (MAD). Angular momentum transport is now mainly
due to the jet and disk winds, and the disk itself is more strongly magnetized ($\beta\sim 1-3$).

In both cases, we find that the pressure anisotropy grows rapidly, up to the point at which it should be limited by the onset of plasma instabilities 
(mainly the mirror instability). At this point, our EMHD model assumes that the pressure anisotropy saturates at the mirror instability threshold, with $\Delta P \approx b^2/2$ 
(for $\beta \gg 1$).
Despite this large pressure anisotropy, the dynamics of the disk remains very similar in the IMHD and EMHD models. In the SANE configuration, the only difference
appears to be a stronger magnetization of the disk in the IMHD simulation (by $\sim 30\%$). In the MAD configuration, there are 
variations of up to $25\%$ in the magnetization of the disk, with only the inner disk being more strongly magnetized in IMHD. 
In both cases, the temperature of the disk does not significantly 
depend on the chosen plasma model, with the ion temperature $\Theta$
consistent with the transformation of potential energy into thermal energy expected in a radiatively inefficient accretion disk. 
The ratio of the surface density to the accretion
rate is also nearly the same in EMHD and IMHD.
In our simulations, the heat flux remains negligible at all times.

In the SANE configuration, we argue that the agreement between the EMHD and IMHD results is largely due to the similarities between 
the shear stresses associated with the magnetic field and pressure anisotropy.
In particular, the shear stress due to pressure anisotropy is identical to the shear stress produced by the magnetic field with the transformation
$\Delta P \rightarrow b^2$ (see Eqs.~\ref{eq:Tmunumag}-\ref{eq:Tmunuvis}). Thus, if $\Delta P \sim b^2/2$ due to the mirror instability,
the impact of the viscous shear stress on the disk dynamics is very similar to that of the B-field.
Practically, we propose that in the innermost regions of a SANE disk, the EMHD results can be derived, in a time-averaged sense at least, 
from ideal MHD results through the simple transformation
\beqn
b^2_{\rm EMHD} &=& (0.7-0.8) b^2_{\rm IMHD}\\
\Delta P_{\rm EMHD} &=& (0.2-0.3) b^2_{\rm IMHD}.
\eeqn
Under that transformation, the evolution equations of the EMHD system are identical to the equations of ideal MHD, up to small corrections
of the order of the magnetic pressure in the energy density and pressure of the fluid. 

We note, however, that this argument is only valid under the restrictive condition that $\beta \gg 1$. 
Accordingly, that rescaling is no longer valid in the MAD configuration.
When $\beta \lesssim 1$, we have $\Delta P \sim P \lesssim b^2/2$. The ratio of the pressure anisotropy 
to the gas pressure is larger than in the SANE case, but the ratio of the pressure anisotropy to the magnetic
pressure is smaller than in the SANE case. The larger value of $\Delta P/P$ in MAD disks, combined with the
qualitative differences between the viscous and magnetic stresses in the presence of a non-negligible 
magnetic pressure, could lead us to believe that non-ideal effects are more important in MAD disks than in SANE
disks. However, in the MAD configuration, the magnetic stresses and pressure dominate over the gas
and viscous stresses, thus reducing the impact of the pressure anisotropy on the dynamics of the disk.
In the end, we find that the second effect is more important, and that there is very little difference between 
EMHD and IMHD MAD disks precisely because the flow dynamics is magnetically dominated ($\beta \leq 1$).

\subsection{Future directions}
\label{sec:future}

The similarities between EMHD and IMHD simulations shown here are a strong indication that accretion disk models neglecting non-ideal effects in the dynamics
of the ions are likely reasonable even for slowly accreting black holes in which the accretion flow is a nearly collisionless plasma. It is, however, important to carefully consider the limitations
of this study. First, all disks studied in this work are relatively compact, and even in the MAD configuration we cannot follow the production of truly unbound outflows
in the disk. The impact of non-ideal effects on disk outflows remains uncertain. In addition, heat-flow driven instabilities such as the magnetothermal 
instability~\citep[MTI]{Balbus2000}
may be active at larger radii in the disk (e.g.~\citealt{Sharma:2008}). 
The growth timescale of the MTI at small radii is likely slower than the accretion timescale, and thus the MTI cannot develop in our 
simulations.

Another important difference between EMHD and IMHD SANE disks is that, in the EMHD simulations, about a third of the accretion power is directly thermalized through the
viscous shear stress due to the pressure anisotropy, instead of through dissipation on small scales at the end of the turbulent cascade driven by the MRI. In our simulations,
this does not appreciably change the ion temperature. However, the different heating mechanisms may very well lead to different ion distribution functions. In addition,
the presence of both viscous and turbulent heating might affect the relative heating of ions and electrons (e.g.~\citealt{Sharma:2007}).
In this work, we solely focus on the dynamics of the disk, 
which is driven by the ions. But modeling the plasma as a single temperature fluid is not sufficient when constructing models for the electromagnetic emission 
from the disk~\citep{Ressler:2016,Sadowski:2016}. 

It is also critical to reiterate that the EMHD model used here is itself an approximation of the complex plasma physics occurring in these disks.  In our calculations, the effects of the mirror instability are particularly important.  The validity of our fluid sub-grid model for collisionless plasmas close to the mirror instability threshold is an active area of research (e.g.,~\citealt{Kunz:2016}).   It is quite possible that surprises will arise that will impact the understanding of sub-Eddington disk dynamics.

Finally, there are interesting outstanding questions about the
connection between our global results and shearing box studies of the
growth and saturation of the MRI.  The latter indicate that the
saturation of the MRI is sensitive to the inclusion of explicit 
(rather than numerical) viscosity and resistivity, and their ratio, 
the magnetic Prandtl number~\citep{Lesur:2007,Fromang:2007}. In our IMHD simulations,
both viscosity and resistivity are set by numerical dissipation. The
numerical viscosity and resistivity both decrease with resolution, but
numerical simulations in ideal MHD have shown that the numerical
magnetic Prandtl number is largely independent of the numerical
resolution~\citep{Guan2009,2009A&A...504..309L,2009A&A...507...19F}, so that effectively all global simulations
of accretion disks in ideal MHD are performed at constant turbulent Prandtl
number.  
On the other hand, the EMHD model introduces a new explicit source of viscosity, which is larger than the numerical viscosity. Should we then expect the MRI to saturate at a very different level in EMHD and IMHD because the former effectively has a larger (albeit anisotropric) microphysical magnetic Prandtl number? If so, how is this consistent with our results which show very little difference between EMHD and IMHD?  Shearing-box simulations with an anisotropic viscosity and an explicit magnetic resistivity may help shed some light on this issue. It may be that including an explicit isotropic viscosity small compared to the anisotropic viscosity is also important in EMHD simulations.

\section*{Acknowledgments}
We thank Matt Kunz and Jono Squire for useful discussions over the course of this project. Support for FF was provided  by NASA through Einstein Postdoctoral Fellowship grant numbered PF4-150122 awarded by the Chandra X-ray Center, which is operated by the Smithsonian Astrophysical Observatory for NASA under contract NAS8-03060. EQ was supported in part by NSF grant AST 13-33612, a Simons Investigator Award from the Simons Foundation, and the David and Lucile Packard Foundation.
CFG was supported
by NSF grant AST-1333612, a Simons Fellowship, and a visiting fellowship at All Souls
College, Oxford. CFG is also grateful to Oxford Astrophysics for a Visiting Professorship
appointment. MC was supported by an Illinois Distinguished
Fellowship from the University of Illinois and by NSF grant AST-1333612. MC also
thanks EQ for a Visiting Scholar appointment at the University of
California, Berkeley, where part of this work was done. We thank Pavan Yalamanichili and the team at {\tt arrayfire.com} for help with performance optimizations in {\tt grim}.
This work was made possible by computing time granted by
the Extreme Science and Engineering Discovery Environment (XSEDE)
through allocation No. TG-PHY160040, supported by NSF Grant
No. ACI-1053575. AT was supported by the Theoretical Astrophysics
Center (TAC) Fellowship and by NSF through XSEDE allocation TG-AST100040.

\bibliographystyle{mnras}
\bibliography{sample}

\appendix

\section{Coordinate transformation}

An important disadvantage of the spherical polar
coordinates commonly used when simulating accretion
onto black holes is that the minimum grid spacing
scales quadratically with resolution, i.e. if the 
number of grid points along each dimension is multiplied
by $n$, the minimum grid spacing is divided by $n^2$. This
is due to the aximuthal grid spacing of the points closest
to the polar axis, which scales as 
$\Delta x_{\rm min} \propto r \times
\Delta \theta \times \Delta \phi$. As the
maximum time step for stable evolution is proportional to
the minimum grid spacing, the cost of 3-dimensional 
simulations scales like $n^5$. 

One way to avoid this issue and recover the same
$n^4$ scaling as for cartesian grids is to use mesh 
refinement and decrease the 
number of cells in the azimuthal direction when close to 
the polar axis. In our simulations, we instead use
coordinate transformations to decrease the resolution 
along $\theta$ close to the polar axis. While our method
leads to more distorted grid cells close to the polar
axis, it has the advantage of avoiding the many 
complications associated with the use of mesh refinement.

Sec.~\ref{sec:coords} describes the structure of
our numerical grid and the main coordinate 
transformations used in our simulations. We define a map
between the coordinates of our numerical grid $(t,a,b,c)$
and the usual Kerr-Schild coordinates $(t,r,\theta,\phi)$
through Eqs.~(\ref{eq:r}-\ref{eq:fr}), plus a map
$\tilde \theta \rightarrow \theta$ described below. 
Eqs.~(\ref{eq:r}-\ref{eq:fr}) focus resolution close to the
black hole, thanks to the exponential radial map. They also
force the use of a map uniform in 
$\tilde \theta$ close to the horizon of the black 
hole, while focusing points in the equatorial regions at
larger radii. A uniform map at small radii is useful to
maximize the minimum grid spacing in the $\theta$ direction,
while focusing points in the equatorial regions provides 
resolution where it is most needed to capture the fastest 
growing mode of the MRI.

The map $\tilde \theta \rightarrow \theta$ is used to avoid
small azimuthal grid spacings at small cylindrical radii near the
poles and follows the approach developed by
\citet{2011MNRAS.418L..79T}. For completeness, we give the details here.
Broadly speaking, this approach transforms the spherical polar 
coordinates into nearly cylindrical coordinates in a small
region within a radius $r_{\rm cyl}$ of the black hole 
center and a distance 
$\rho_{\rm cyl} \sim r_{\rm cyl} \theta_{\rm cyl}$ 
of the polar axis. More precisely, for 
$0\leq \theta \leq \pi/2$, the map is defined as
\beq
\sin(\theta) = S_{\rm max}\left(\sin(\tilde \theta),f_2(r,\tilde \theta),\frac{r_{\rm cyl}+r_{\rm in}}{2r} df\right),
\label{eq:cylmap}
\eeq
with $r_{\rm in}$ the radius of the inner boundary.

The function $S_{\rm max}$ is defined as
\beq
S_{\rm max}(x_0,x_1,dx) = x_1 - dx F\left(\frac{x_1-x_0}{dx}\right),
\eeq
with $F(x)=0$ for $x<-1$, $F(x)=x$ for $x>1$, and
\beqn
2F(x) &=& x+1+
\frac{1}{160\pi} \sin\left(\frac{5\pi(x+1)}{2}\right) + \frac{5}{96\pi} \sin\left(\frac{3\pi(x+1)}{2}\right)
\nonumber\\
&& - \frac{35}{16\pi} \sin\left(\frac{\pi(x+1)}{2}\right)
\eeqn
for $-1 \leq x \leq 1$.
Practically, $S_{\rm max}$ returns the maximum 
of $(x_0,x_1)$ if
$x_0 \gg x_1$ or $x_0 \ll x_1$, and smoothly transitions
between $x_0$ and $x_1$ when $x_0 \sim x_1$. 
Eq.~\ref{eq:cylmap} is thus a smooth transition between
$\theta = \tilde \theta$ and $\sin(\theta) = f_2(r,\tilde \theta)$. The latter is the `cylindrified' coordinate,
defined as
\beq
f_2(r,\tilde\theta) = S\left(
\frac{r_{\rm cyl}}{r}\sin(\tilde\theta),
\sin \left(\theta_2(r,\tilde\theta)\right),
|\sin{(\theta_2(r,0)})|,
r-r_{\rm cyl}
\right),
\eeq
where $S(x_0,x_1,dx,A)=S_{\rm max}(x_0,x_1,dx)$ if $A>0$ and
$S(x_0,x_1,dx,A)=-S_{\rm max}(-x_0,-x_1,dx)$ for $A<0$. The
angle $\theta_2$ is
\beqn
\theta_2(r,\tilde\theta) &=& \theta_1(r)
+ \left(\pi/2 - \theta_1(r)\right) \frac{\tilde \theta-\theta_{\rm cyl}}{\pi/2-\theta_{\rm cyl}},\\
\sin{\left(\theta_1(r)\right)} &=&
\frac{r_{\rm cyl}}{r}
\sin{(\theta_{\rm cyl})}.
\eeqn

In the limit 
$\sin (\theta) = f_2 = \frac{r_{\rm cyl}}{r} \sin{(\tilde \theta)}$, lines
of constant $\tilde \theta$ are vertical lines at a constant
distance $\rho = r_{\rm cyl} \sin{(\tilde \theta)}$ from the polar
axis. The smoothing 
functions are chosen so that 
$f_2 \approx \sin{\tilde\theta}$ for 
$\tilde \theta \gg \theta_{\rm cyl}$, while
$f_2 \approx \frac{r_{\rm cyl}}{r} \sin{(\tilde \theta)}$ for
$\tilde \theta \lesssim \theta_{\rm cyl}$.

The smoothing width $\frac{r_{\rm cyl}+r_{\rm in}}{2r} df$, 
on the other hand, provides the transition
between `cylindrified' coordinates $f_2$ and `standard'
coordinates $\sin{(\tilde\theta)}$ around 
$r\sim r_{\rm cyl}$. We use
\beq
df = |f_2(\frac{r_{\rm cyl}+r_{\rm in}}{2},\tilde\theta) - \sin{(\tilde\theta)}|.
\eeq

In practice, we choose 
$\theta_{\rm cyl} \sim \pi/N_\theta$, with
$N_\theta$ the number of grid cells in the polar direction.
The map $\tilde \theta \rightarrow \theta$ thus only 
affects the grid cells closest to the polar axis.
The parameter $r_{\rm cyl}$ is resolution dependent, and
chosen to maximize the smallest grid spacing in our
computational domain (we use $r_{\rm cyl} = 8 r_{\rm in}$
in the high-resolution simulations of SANE disks presented
in this work). The map for $\pi/2 < \theta \leq \pi$
is obtained by requiring symmetry across
the equatorial plane.

\label{lastpage}

\end{document}